\documentclass[aps,prl,reprint,superscriptaddress,amsmath]{revtex4-1}
\usepackage{lipsum}
\usepackage{graphicx}
\usepackage{subfigure}
\usepackage{textcomp}
\usepackage{setspace}
\begin{document}
\title{Pressure-induced structural phase transition of vanadium:\\
  A revisit from the perspective of ensemble theory}
\author{Bo-Yuan Ning}
\affiliation{Institute of Modern Physics, Fudan University, Shanghai, 200433, China}
\affiliation{Applied Ion Beam Physics Laboratory, Fudan University, Shanghai, 200433,
China}
\affiliation{Department of Materials Science and Engineering,
  Southern University of Science and Technology, Shenzhen 518055, China}
\author{Xi-Jing Ning}
\email{xjning@fudan.edu.cn}
\affiliation{Institute of Modern Physics, Fudan University, Shanghai, 200433, China}
\affiliation{Applied Ion Beam Physics Laboratory, Fudan University, Shanghai, 200433,
China}

\date{\today}

\begin{abstract}
  For realistic crystals, the free energy strictly formulated in ensemble theory can hardly be obtained because of the difficulty in solving the high-dimension integral of the partition function, the dilemma of which makes it even a doubt if the rigorous ensemble theory is applicable to phase transitions of condensed matters.
  In the present work, the partition function of crystal vanadium under compression up to $320$ GPa at room temperature is solved by an approach developed very recently,
  and the derived equation of state is in a good agreement with all the experimental measurements,
  especially the latest one covering the widest pressure range up to 300 GPa.
  Furthermore, the derived Gibbs free energy proves the very argument to understand most of the experiments reported in the past decade on the pressure-induced phase transition, and, especially,
  a novel phase transition sequence concerning three different phases observed very recently and
  the measured angles of two phases agree with our theoretical results excellently.
\end{abstract}

\maketitle

The structural phase transition of crystal vanadium (V) under high pressure at room temperature attracts a long-time interests.
Although it had been already a consensus that the BCC phase is highly stable\cite{vstable1} and early experiments affirmed this stability up to $220$ GPa\cite{vstable2},
theoretical works based on calculations of phonon mode softening and trigonal shear elastic instability predicted a likely structural transformation within a range from $130$ GPa\cite{Suzuki2002} to $200$ GPa\cite{Landa_2006, landa2006}, which inspired an experiment\cite{Ding} in year 2007 showing that a phase transition does take place at about $63$ GPa. Specifically, the BCC phase transits into a rhombohedral (RH) structure with the RH angle $\alpha>\alpha_{\text{BCC}}$($109.47^\circ$) denoted as RH$_1$ to distinguish from a similar structure RH$_2$ with $\alpha<\alpha_{\text{BCC}}$.

The above mentioned phase transition was further confirmed by later experiments \cite{jenei, vshock, vphase, wang2021, vexp2016}, and the relevant theoretical works, based on either approximate calculations of free energy\cite{wuqiang16,qiu2008} or \emph{ab initio} lattice dynamics\cite{luo07,Verma2007, lee1, lee2,kohnanomaly2018}, reached a qualitative agreement that the transition pressure ($P_c$) would be $60\sim90$ GPa for BCC$\rightarrow$RH$_1$, and predicted that $P_c$ for RH$_1\rightarrow$ RH$_2$ and RH$_2\rightarrow$ BCC are around $120$ and $250$ GPa respectively.
Nevertheless, two experiments reported in year 2021\cite{akahama2021,mcmahon2021} exhibited different results.
Akahama \emph{et al}.\cite{akahama2021} compressed foil V up to $300$ GPa at room temperature and found that BCC lattice is a stable phase until the RH$_2$ and BCC phase coexisting at pressure ($P$) larger than $242$ GPa, 
while the RH$_1$ phase is a metastable phase caused by nonhydrostatic pressure effects.
Stevenson \emph{et al}.\cite{mcmahon2021}, on the other hand,
observed that the BCC lattice transform into RH$_2$ when $P>40$ GPa and then turns back to BCC until RH$_1$ emerges for $P>100$ GPa.
It is noticeable that the measured $\alpha$ by Stevenson \emph{et al}. for RH$_1$ (or RH$_2$) is $109.54^\circ$ (or $109.35^\circ$),
which is significantly different from the theoretical result, $110.5^\circ$ (or $108.5^\circ$) for RH$_1$ (or RH$_2$).

In the viewpoint of statistical mechanics, all the discrepancies stated above as well as others mentioned in literatures\cite{pfconden,mce,fe} should be settled down as long as the partition function (PF) can be obtained to produce the free energy (FE). Unfortunately, the exact solution to the PF of condensed matters is almost impossible because of the $3N$-fold configurational integral, and various approximations\cite{liu2020}, such as the one in Ref.\cite{qiu2008}, were developed to calculate the FE without knowledge of the PF.
As expected, those approximated methods provided lots of interesting information on the phase transitions of condensed matters, while may not address all the issues substantially. 

Very recently, we put forward a direct integral approach (DIA) to the PF of condensed state systems with ultrahigh efficiency and precision\cite{nby,lyp,glc1,glc2}, and has been successfully applied to reproduce the equation of state (EOS) for solid copper\cite{nby}, argon\cite{glc1} and $2$-D materials\cite{lyp} obtained from experiments or molecular dynamics simulations.
Compared with phonon model based on harmonic or quasi-harmonic approximations,
which is currently applied to produce EOS,
DIA is applicable to much wider realm with much higher precision\cite{glc2}.
In the present work, DIA is used to compute the PF of crystal V with various phases and the derived Gibbs FE is applied to investigate the phase transitions induced by pressure at room temperature. 

For a crystal containing $N$ atoms confined within volume $\mathcal{V}$ at temperature $T$,
the atoms are regarded as $N$ point particles of the atomic mass $m$ with Cartesian coordinate $\mathbf{q}^N=\{\mathbf q_1,\mathbf q_1,\ldots\mathbf q_N\}$,
and the total potential energy, $U(\mathbf{q}^N)$,
as the function of $\mathbf{q}^N$ is computed by quantum mechanics,
i.e., for a given set of $\mathbf{q}^N$,
the total potential energy $U(\mathbf{q}^N)$ concerned with the motions of electrons in the field of the nucleus fixed at the lattice sites is calculated by quantum mechanics.
With knowledge of $U(\mathbf{q}^N)$, the PF of the system reads
\begin{equation}
  \begin{split}
  \label{eq:1}
  \mathcal{Z}&=\frac{1}{N!}\left(\frac{2\pi m}{\beta h^2}\right)^{\frac{3}{2}N}\int d\textbf{q}^N\exp[-\beta U(\textbf{q}^N)]\\
  &=\frac{1}{N!}\left(\frac{2\pi m}{\beta h^2}\right)^{\frac{3}{2}N}\mathcal Q,
  \end{split}
\end{equation}
where $h$ is the Planck constant and $\beta=1/k_BT$ with $k_B$ the Boltzmann constant. 
If the configurational integral $\mathcal{Q}=\int d\textbf{q}^N\exp[-\beta U(\textbf{q}^N)]$ is solved, then the pressure $P$ and the Gibbs FE $G$ can be computed by 
\begin{eqnarray}
  \label{eq:3}
  P&=&\frac{1}{\beta}\frac{\partial\ln\mathcal{Q}}{\partial\mathcal{V}}, \\
  \label{eq:4}
  G&=&-\frac{1}{\beta}\ln[\frac{1}{N!}\left(\frac{2\pi m}{\beta h^2}\right)^{\frac{3}{2}N}]-\frac{1}{\beta}\ln\mathcal{Q}+P\mathcal{V}.
\end{eqnarray}
In this way, the contributions from both the electrons and nucleus are included in the calculations.

\begin{figure}
  \centering
  \includegraphics[width=2.9in,height=1.8in]{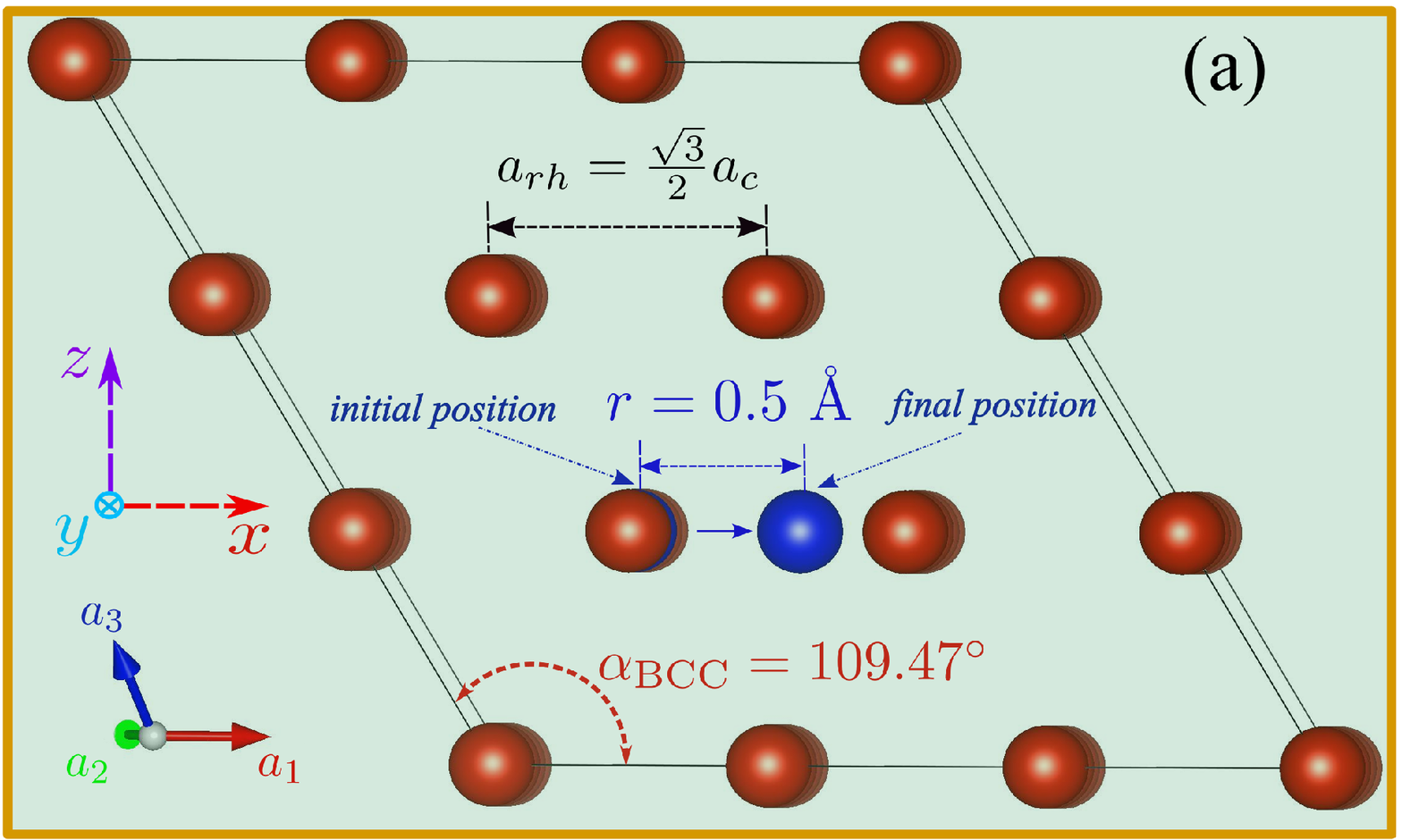}
  \includegraphics[width=3.2in,height=2.4in]{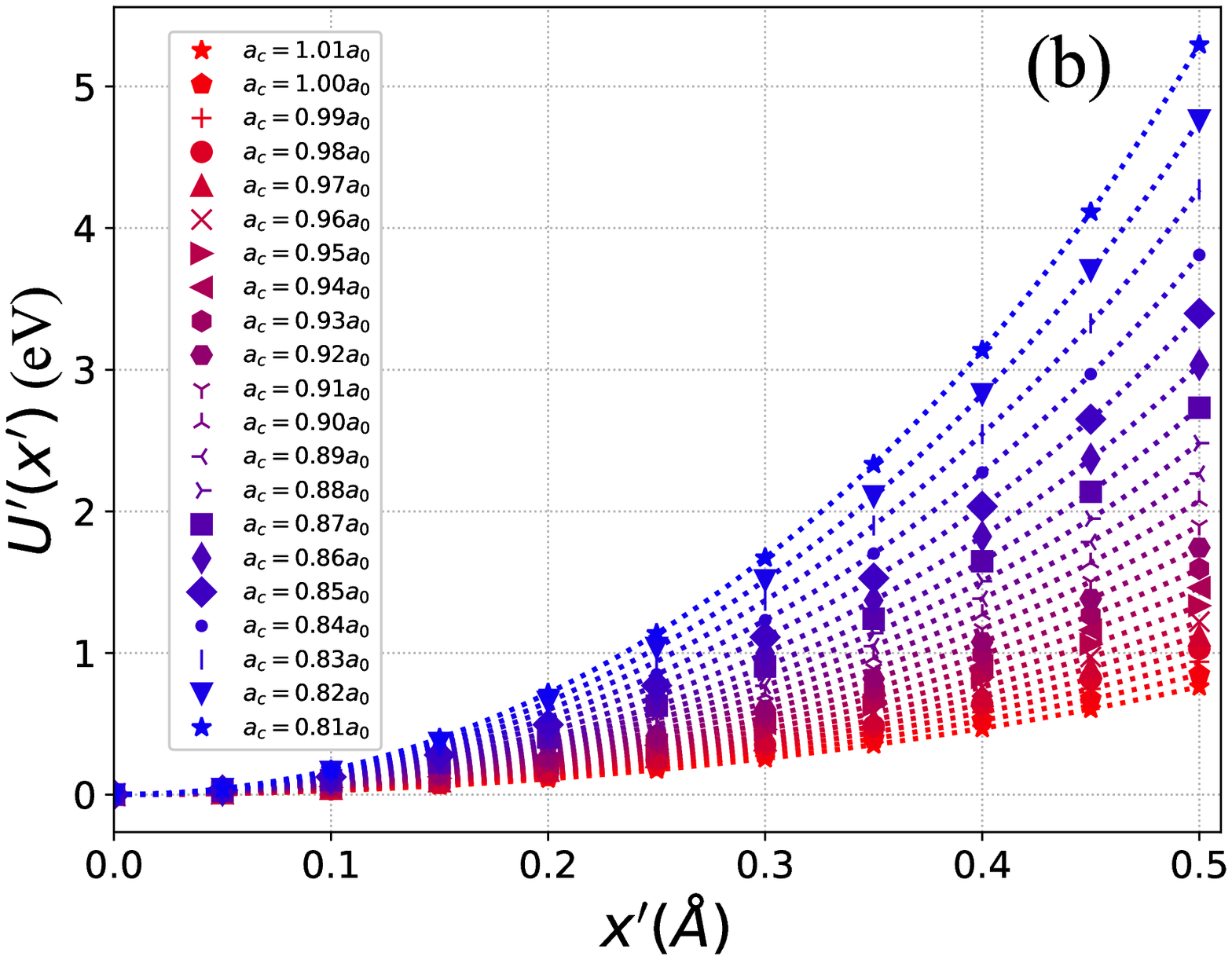}
  \caption{(Color Online) (a) Schematic of DIA to the BCC structure in a RH supercell and (b) the calculated $U'(x')$ for different supercell volumes with $a_c$ changed from $0.81$ to $1.01a_0$ ($a_0=3.0${\AA}).}
  \label{fig:1}
\end{figure}

According to our proposed DIA\cite{nby}, for a single-component crystal with $N$ atoms placed in their lattice sites $\mathbf Q^N$ and with the total potential energy $U_0(\mathbf{Q}^N)$, we firstly introduce a transformation, 
\begin{equation}
  \label{eq:2}
  \mathbf q'^N=\mathbf q^N-\mathbf Q^N,\ U'(\mathbf q'^N)=U(\mathbf q'^N)-U_0(\mathbf Q^N), 
\end{equation}
where $\mathbf q'^N$ represents the displacements of atoms away from their lattice positions, and then the configurational integral can be expressed in a one-fold integral,
\begin{equation}
  \label{eq:5}
  \mathcal{Q}=e^{-\beta U_0}\left[\int e^{-\beta U'(q'_{i_{x,y,z}})}dq'_{i_{x,y,z}}\right]^{3N}=e^{-\beta U_0}\mathcal{L}^{3N},
\end{equation}
where $q'_{i_{x,y,z}}$ denotes the distance of the $i$th atom moving along the $x$ (or $y$, $z$) direction relative to its lattice site while the other two degrees of freedom of the atom and all the other atoms are kept fixed.
As shown in Fig.\ref{fig:1}, we take BCC structure of V placed in a $3\times3\times3$ RH supercell as an example to illustrate the implementation of the DIA.
The basis vectors of the primitive cell are set as $\mathbf{a}_1=a_{\text{rh}}\cdot(1, 0, 0)$, $\mathbf{a}_2=a_{\text{rh}}\cdot(\cos\alpha, \sin\alpha, 0)$, $\mathbf{a}_3=a_{\text{rh}}\cdot(\cos\alpha, \frac{\cos\alpha-\cos^2\alpha}{\sin\alpha}, \frac{\sqrt{1-3\cos^2\alpha+2\cos^3\alpha}}{\sin\alpha})$ and the volume of the cell equals to $\displaystyle a_{\text{rh}}^3\sqrt{1-3\cos^2\alpha+2\cos^3\alpha}$, where $a_{\text{rh}}=\sqrt{3}/2a_c$ with $a_c$ for the lattice constant of a cubic primitive cell.
According to Eq.(\ref{eq:5}), an arbitrary atom is selected and moved $0.5$ {\AA} by a step of $0.05$ {\AA} along the direction of $\mathbf{a}_1$ as shown in Fig.\ref{fig:1}(a) with the initial and final positions of the atom colored in blue.
During the movement, the total potential energy is computed by the density functional theory (DFT) at every steps and the spline interpolation algorithm\cite{spl1,*spl2} is used to smooth the $U'(x')$ curves, which are shown in Fig.\ref{fig:1}(b).

The DFT calculations is performed in Vienna Ab initio Simulation Package\cite{VASP1,VASP2} with the projector-augmented wave formalism\cite{paw1,paw2}, and the general gradient approximation of the Perdew-Burke-Ernzerhf parametrizations\cite{pbe} is adopted for the exchange-correlation functional with 13 valence electrons ($3s^2p^6d^34s^2$) considered\cite{gga1}.
A $\Gamma$-centered $9\times9\times9$ uniform $k$-mesh grid is set to sample the Brillouin zone by the Monkhorst-Pack scheme\cite{monkhorst} and the tetrahedron method with Bl\"ochl corrections is used to determine the electron orbital partial occupancy, together with $342.7749$ eV set as the cut-off energy of the plane-wave basis, $1\times10^{-6}$ eV as the convergence energy criterion of the electron self-consistent computations.

\begin{figure}
  \centering
  \includegraphics[width=3.2in,height=2.5in]{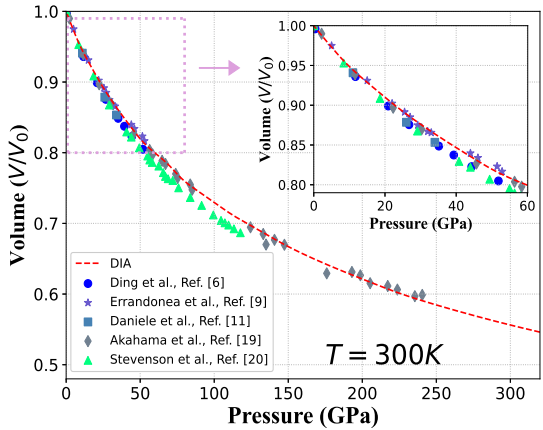}
  \caption{(Color Online) Isothermal $P$-$\mathcal{V}$ curve of the BCC phase at $300$K from the PF (red dash line) and the experimental data denoted in different colored symbols. The inset shows the detailed comparisons in the low-pressure zone.}
  \label{fig:2}
\end{figure}

As shown in Fig.\ref{fig:2},
the isothermal $P$-$\mathcal V$ curve derived from the PF via Eq.(\ref{eq:3}) coincides very well with the experiments,
which can be seen from the relative difference (RD) between the theoretical volume $(V/V_0)_{\text{PF}}$ and the experimental one $(V/V_0)_{\text{exp}}$,
defined as RD$=\frac{|(V/V_0)_{\text{PF}}-(V/V_0)_{\text{exp}}|}{(V/V_0)_{\text{exp}}}$,
where $V_{0_{\text{PF}}}$ and $V_{0_{\text{exp}}}$ are the theoretical and experimental atomic volumes under ambient conditions (room temperature and one atmospheric pressure).
In the low-pressure zone ($P<20$ GPa), the average RD for all the shown experiments is about $0.25\%$,
which can be seldom achieved in common theoretical work without using empirical data or empirical EOS.
In the higher-pressure zone ($P>20$ GPa), the RD for all the experiments except for the one of Ref.\cite{akahama2021} get a little larger with the pressure increasing,
which may be attributed to the difficulty in precise measurements of higher pressure in considerations of the fact that the results of different experiments diverge larger with increases of the pressure (see the experimental points shown in the inset of Fig.\ref{fig:2}).
It is worthwhile to see that the RD for one of the latest experiments\cite{akahama2021} (grey diamonds shown in Fig.\ref{fig:2}),
covering the widest pressure range, keeps to be smaller than $0.7\%$ except for three pressure points at $135$, $176$ and $240$ GPa.
Such an excellent agreement enables us to theoretically determine the lattice constant under exactly given pressure and temperature.
As listed in Table.\ref{tab:1}, the RD between the theoretical $a_c$ and the experimental ones is smaller than $0.73\%$,
exhibiting a good agreement.

It should be pointed out that the implementation of DIA excludes all the artificial dependence of adjustable parameters,
empirical EOS or experimental data, 
and accordingly,
the good agreements between the theory and the experiments for the EOS and the lattice constant strongly indicate that
the ensemble theory is the very approach to understand the thermodynamics properties of condensed matters.

\begin{table}[htbp]
  \centering
  \caption{The lattice constant $a_c$ of BCC phase at ambient conditions calculated by DIA and from experiments fitting empirical EOSs, Birch-Murnaghan (B-M), Vinet and AP2.}
  \label{tab:1}
  \begin{ruledtabular}
    \begin{tabular}{l|ccccc}
     & \small DIA & \small Ref.\cite{vphase} & \small Ref.\cite{vexp2016} & Ref.\cite{akahama2021} & \small Ref.\cite{mcmahon2021}\\
      \hline
      $a_c$ ({\AA})& 3.008 & 3.029 & 3.03 & 3.027 & 3.023 \\
      \small Deviations & \small / & $0.693\%$ &  $0.726\%$ & $0.627\%$ & $0.496\%$ \\
      \small EOS & \small / & \small B-M & \small \emph{no report} & \small Vinet & \small AP2
    \end{tabular}
  \end{ruledtabular}
\end{table}

\begin{figure}
  \centering
  \includegraphics[width=3.2in,height=2.4in]{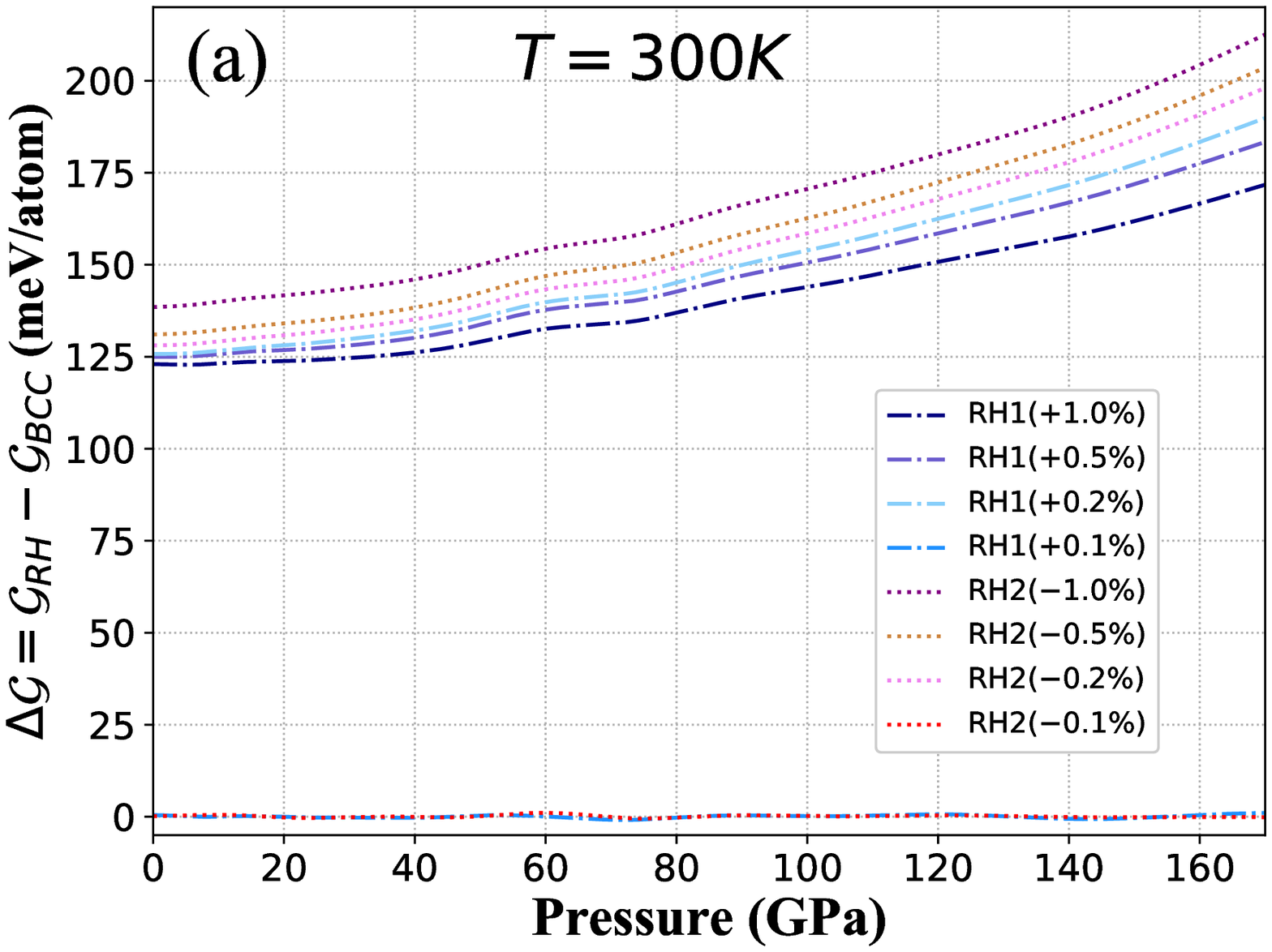}
  \includegraphics[width=3.2in,height=2.4in]{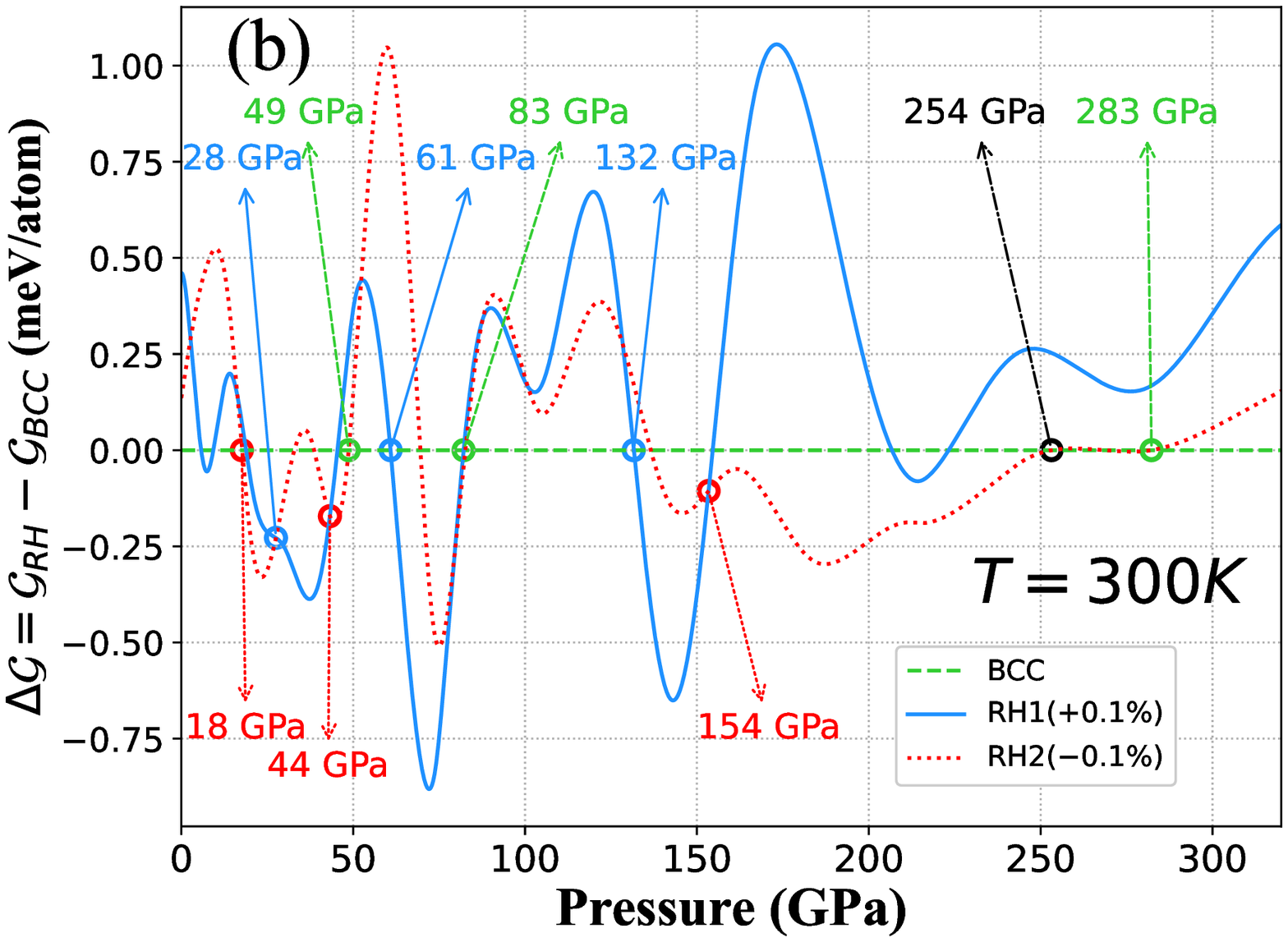}
  \caption{(Color Online) (a) Gibbs FE of RH phases relative to BCC phase at room temperature up to $170$ GPa and (b) the data for RH phases with $\pm 0.1\%\cdot\alpha_{\text{BCC}}$ for clarity.}
  \label{fig:3}
\end{figure}

To consider the transitions of crystal V from the BCC to RH phases with the angle $\alpha$ different from $\alpha_{\text{BCC}}$,
the PF of four RH$_1$ and RH$_2$ phases with the angle deviation, $\Delta=(\alpha_{\text{RH}}-\alpha_{\text{BCC}})/\alpha_{\text{BCC}}$,
$\pm0.1\%$, $\pm0.2\%$, $\pm0.5\%$ and $\pm1\%$ is solved by the DIA in the same way as described above (see potential-energy curves of the RH phases in supplementary material), and,
the derived Gibbs FE differences between the RH and BCC per atom, $\mathcal G_{\text{RH}}-\mathcal G_{\text{BCC}}$, are shown in Fig.\ref{fig:3}(a).
When the amplitude of the angle deviation, $|\Delta|$, equals to $0.1\%$, the Gibbs FE of both RH$_1$ and RH$_2$ is close to $\mathcal G_{\text{BCC}}$,
and increases abruptly to be larger than $\mathcal G_{\text{BCC}}$ by more than $120$ meV/atom with the $|\Delta|$ increased by only $0.1\%$,
while further increasing the angle deviation leads to a little changes of $\mathcal G_{\text{RH}}$.
These results show clearly, based on the Gibbs FE criterion,
that the phase transition with the angle equal to or larger than $0.2\%$ cannot take place
unless the deviation being smaller, and furthermore, Fig.\ref{fig:3}(b) indicates that the two phases,
RH$_1$ with a deviation of $0.1\%$ and RH$_2$ with a deviation of $-0.1\%$,
would emerge when the pressure is larger than $20$ GPa.
It is interesting to note that the phase transition was very recently examined by Stevenson et al.\cite{mcmahon2021}
who concluded that two kinds of phase transitions, BCC$\rightarrow$ RH$_1$ and BCC $\rightarrow$ RH$_2$,
indeed take place,
qualitatively coinciding with the previous experiments or theories, 
and the determined RH angle for RH$_1$ (or RH$_2$) is $109.54^\circ$ (or $109.35^\circ$), corresponding to $|\Delta|\sim0.1\%$ as our theoretical results,
which is a little smaller than the one, $109.65^\circ$ for RH$_1$, measured in previous experiment\cite{Ding,jenei} 
where RH$_2$ was not observed,
but significantly different from the theoretically predicted $110.5^{\circ}$ (or $108.5^\circ$)\cite{luo07,lee1,Verma2007,qiu2008,wuqiang16}, corresponding to $|\Delta|\sim 1\%$.
Since the unique difference between BCC and RH$_1$ (or RH$_2$) phase is the RH angle deviation of about only $0.1\%$,
it should be quite a challenge for experimental observations of the subtle difference,
which may be the reason
why early literatures reported no phase transitions for V induced by pressure less than $220$ GPa\cite{vstable2}.

Now we make a detailed comparison between our theoretical results and the experimental observations.
As shown in Fig.\ref{fig:3}(b),
the Gibbs FE of the RH$_1$ and RH$_2$ differ from that of BCC less than $1$ meV,
which is much smaller than the average kinetic energy, $\sim30$ meV,
of a thermal atom at room temperature,
and it seems that RH$_1$ or RH$_2$ is just the thermal fluctuation from the BCC structure instead of a phase transition.
Actually, the very ``driving force'' accounting for the phase transition under given pressure and temperature is not the thermal kinetic energy (or internal energy),
but instead, the Gibbs FE that determines the probability for a given phase existence.
According to ensemble theory,
the relative probability for RH with respective to BCC equals to $e^{(G_\text{BCC}-G_\text{RH})/k_BT}$,
where $G_{\text{BCC}}=N\mathcal{G}_{\text{BCC}}$, $G_{\text{RH}}=N\mathcal{G}_{\text{RH}}$,
and $N$ the number of the atoms in a piece of macro bulk crystal is on an order of $10^{23}$.
Thus, our calculation results indicate clearly that the
phase transition BCC $\rightarrow$ RH$_2$ must take place at room temperature when the pressure is larger than $18$ GPa, and then RH$_1$ emerges at $28$ GPa.
This RH$_1$ transition was observed by Jenei et al.\cite{jenei} at $32$ GPa and by Daniele et al.\cite{vexp2016} at $35$ GPa, respectively.
As the pressure increasing,
our results show that RH$_2$ would emerge at $\sim44$ GPa,
then transforms back to BCC for the pressure larger than $50$ GPa until RH$_1$ emerges at $61$ GPa,
displaying a transition process,
RH$_2\rightarrow$ BCC $\rightarrow$ RH$_1$,
with the pressure increased.
Experimentally,
Akahama et al.\cite{akahama2021} observed a RH phase comes to existence from $45$ GPa,
and the BCC $\rightarrow$ RH$_1$ transition was reported at $63$ GPa in Ref.\cite{Ding},
$60.5$ GPa in Ref.\cite{vshock} and $64$ GPa in Ref.\cite{vphase}, 
which are all in a good agreement with our results,
in considerations of the deviations among the measured pressures being of $\sim 2$ GPa.
It is noticeable that
the RH$_2\rightarrow$ BCC $\rightarrow$ RH$_1$ transition sequence,
neither observed in earlier experiments nor proposed by previous computations,
was observed in the one of the latest experiments\cite{mcmahon2021},
and the measured transition pressure of RH$_2$, $40$ GPa,
coincides with our theoretical one, $44$ GPa.
When the pressure is increased larger than $132$ GPa,
our results indicate that RH$_1$ would exist until RH$_2$ transition occurs at $154$ GPa
and the RH$_2$ phase keeps stable with the pressure up to $254$ GPa,
over which the BCC may coexist with RH$_2$ until the pressure increases up to $280$ GPa
when BCC becomes the most stable for higher pressures.
It is very interesting to note that
this coexistence phase of RH$_2$ and BCC,
never reported in previous literatures,
was observed by Akahama et al.\cite{akahama2021} very recently at $242$ GPa.

The above discussions show that most results of our calculations coincide well with the experiments,
especially the EOS, the angle deviations of the RH and some transition pressures,
such as $30$, $40$, $60$, $250$ GPa for
BCC $\rightarrow$ RH$_1$, RH$_2$ existence, BCC $\rightarrow$ $RH_1$ and coexistence of RH$_2$ and BCC, respectively.
Based on these facts,
it should be reasonable to expect that
the disagreements between our results and the experimental observations,
such as the theoretical phase transition under $\sim20$ GPa,
which was not reported in all the experiments,
may be settled down by future experiments.

In conclusion,
the rigorous ensemble theory is applied, for the first time,
to investigate phase transitions of crystal V via DIA solving the PF without any artificial tunable parameters or empirical EOS,
achieving very good agreements with experimental measurements and observations,
and showing that the theory is substantial to describe phase transitions of condensed matters.
The theoretical approach of this work may find its vast applications in the field to predict parameter-free EOS and phase behaviors of condensed matters under extreme conditions.

\section{acknowledgement}
BYN is grateful to A. Daniele, M. McMahon and Y. Akahama for their kindly providing the raw experimental data in Refs.\cite{vexp2016}, \cite{akahama2021} and \cite{mcmahon2021}, respectively. Part of the computational tasks was conducted in HPC platform supported by The Major Science and Technology Infrastructure Project of Material Genome Big-science Facilities Platform supported by Municipal Development and Reform Commission of Shenzhen.

\begin{thebibliography}{37}%
\makeatletter
\providecommand \@ifxundefined [1]{%
 \@ifx{#1\undefined}
}%
\providecommand \@ifnum [1]{%
 \ifnum #1\expandafter \@firstoftwo
 \else \expandafter \@secondoftwo
 \fi
}%
\providecommand \@ifx [1]{%
 \ifx #1\expandafter \@firstoftwo
 \else \expandafter \@secondoftwo
 \fi
}%
\providecommand \natexlab [1]{#1}%
\providecommand \enquote  [1]{``#1''}%
\providecommand \bibnamefont  [1]{#1}%
\providecommand \bibfnamefont [1]{#1}%
\providecommand \citenamefont [1]{#1}%
\providecommand \href@noop [0]{\@secondoftwo}%
\providecommand \href [0]{\begingroup \@sanitize@url \@href}%
\providecommand \@href[1]{\@@startlink{#1}\@@href}%
\providecommand \@@href[1]{\endgroup#1\@@endlink}%
\providecommand \@sanitize@url [0]{\catcode `\\12\catcode `\$12\catcode
  `\&12\catcode `\#12\catcode `\^12\catcode `\_12\catcode `\%12\relax}%
\providecommand \@@startlink[1]{}%
\providecommand \@@endlink[0]{}%
\providecommand \url  [0]{\begingroup\@sanitize@url \@url }%
\providecommand \@url [1]{\endgroup\@href {#1}{\urlprefix }}%
\providecommand \urlprefix  [0]{URL }%
\providecommand \Eprint [0]{\href }%
\providecommand \doibase [0]{http://dx.doi.org/}%
\providecommand \selectlanguage [0]{\@gobble}%
\providecommand \bibinfo  [0]{\@secondoftwo}%
\providecommand \bibfield  [0]{\@secondoftwo}%
\providecommand \translation [1]{[#1]}%
\providecommand \BibitemOpen [0]{}%
\providecommand \bibitemStop [0]{}%
\providecommand \bibitemNoStop [0]{.\EOS\space}%
\providecommand \EOS [0]{\spacefactor3000\relax}%
\providecommand \BibitemShut  [1]{\csname bibitem#1\endcsname}%
\let\auto@bib@innerbib\@empty
\bibitem [{\citenamefont {Takemura}(2000)}]{vstable1}%
  \BibitemOpen
  \bibfield  {author} {\bibinfo {author} {\bibfnamefont {K.}~\bibnamefont
  {Takemura}},\ }\href@noop {} {\emph {\bibinfo {title} {Science and Technology
  of High Pressure}}},\ edited by\ \bibinfo {editor} {\bibfnamefont {M.~H.}\
  \bibnamefont {Manghnani}}, \bibinfo {editor} {\bibfnamefont {W.~J.}\
  \bibnamefont {Nellis}}, \ and\ \bibinfo {editor} {\bibfnamefont {M.~F.}\
  \bibnamefont {Nicol}}\ (\bibinfo  {publisher} {Universities Press, Hyderabad,
  India},\ \bibinfo {year} {2000})\ p.\ \bibinfo {pages} {443}\BibitemShut
  {NoStop}%
\bibitem [{\citenamefont {Nakamoto}\ \emph {et~al.}(2005)\citenamefont
  {Nakamoto}, \citenamefont {Takemura}, \citenamefont {Ishizuka}, \citenamefont
  {Shimizu},\ and\ \citenamefont {Kikegawa}}]{vstable2}%
  \BibitemOpen
  \bibfield  {author} {\bibinfo {author} {\bibfnamefont {Y.}~\bibnamefont
  {Nakamoto}}, \bibinfo {author} {\bibfnamefont {K.}~\bibnamefont {Takemura}},
  \bibinfo {author} {\bibfnamefont {M.}~\bibnamefont {Ishizuka}}, \bibinfo
  {author} {\bibfnamefont {K.}~\bibnamefont {Shimizu}}, \ and\ \bibinfo
  {author} {\bibfnamefont {T.}~\bibnamefont {Kikegawa}},\ }in\ \href@noop {}
  {\emph {\bibinfo {booktitle} {Joint $20^{th}$ AIRAPT - $43^{th}$ EHPRG,
  Karlsruhe/Germany}}}\ (\bibinfo {year} {2005})\BibitemShut {NoStop}%
\bibitem [{\citenamefont {Suzuki}\ and\ \citenamefont
  {Otani}(2002)}]{Suzuki2002}%
  \BibitemOpen
  \bibfield  {author} {\bibinfo {author} {\bibfnamefont {N.}~\bibnamefont
  {Suzuki}}\ and\ \bibinfo {author} {\bibfnamefont {M.}~\bibnamefont {Otani}},\
  }\href@noop {} {\bibfield  {journal} {\bibinfo  {journal} {J. Phys.: Condens.
  Matter}\ }\textbf {\bibinfo {volume} {14}},\ \bibinfo {pages} {10869}
  (\bibinfo {year} {2002})}\BibitemShut {NoStop}%
\bibitem [{\citenamefont {Landa}\ \emph
  {et~al.}(2006{\natexlab{a}})\citenamefont {Landa}, \citenamefont {Klepeis},
  \citenamefont {Söderlind}, \citenamefont {Naumov}, \citenamefont
  {Velikokhatnyi}, \citenamefont {Vitos},\ and\ \citenamefont
  {Ruban}}]{Landa_2006}%
  \BibitemOpen
  \bibfield  {author} {\bibinfo {author} {\bibfnamefont {A.}~\bibnamefont
  {Landa}}, \bibinfo {author} {\bibfnamefont {J.}~\bibnamefont {Klepeis}},
  \bibinfo {author} {\bibfnamefont {P.}~\bibnamefont {Söderlind}}, \bibinfo
  {author} {\bibfnamefont {I.}~\bibnamefont {Naumov}}, \bibinfo {author}
  {\bibfnamefont {O.}~\bibnamefont {Velikokhatnyi}}, \bibinfo {author}
  {\bibfnamefont {L.}~\bibnamefont {Vitos}}, \ and\ \bibinfo {author}
  {\bibfnamefont {A.}~\bibnamefont {Ruban}},\ }\href@noop {} {\bibfield
  {journal} {\bibinfo  {journal} {Journal of Physics: Condensed Matter}\
  }\textbf {\bibinfo {volume} {18}},\ \bibinfo {pages} {5079} (\bibinfo {year}
  {2006}{\natexlab{a}})}\BibitemShut {NoStop}%
\bibitem [{\citenamefont {Landa}\ \emph
  {et~al.}(2006{\natexlab{b}})\citenamefont {Landa}, \citenamefont {Klepeis},
  \citenamefont {S\"{o}derlind}, \citenamefont {Naumov},\ and\ \citenamefont
  {Velikokhatnyi}}]{landa2006}%
  \BibitemOpen
  \bibfield  {author} {\bibinfo {author} {\bibfnamefont {A.}~\bibnamefont
  {Landa}}, \bibinfo {author} {\bibfnamefont {J.}~\bibnamefont {Klepeis}},
  \bibinfo {author} {\bibfnamefont {P.}~\bibnamefont {S\"{o}derlind}}, \bibinfo
  {author} {\bibfnamefont {I.}~\bibnamefont {Naumov}}, \ and\ \bibinfo {author}
  {\bibfnamefont {O.}~\bibnamefont {Velikokhatnyi}},\ }\href@noop {} {\bibfield
   {journal} {\bibinfo  {journal} {J. Phys. Chem. Solids}\ }\textbf {\bibinfo
  {volume} {67}},\ \bibinfo {pages} {2056 } (\bibinfo {year}
  {2006}{\natexlab{b}})}\BibitemShut {NoStop}%
\bibitem [{\citenamefont {Ding}\ \emph {et~al.}(2007)\citenamefont {Ding},
  \citenamefont {Ahuja}, \citenamefont {Shu}, \citenamefont {Chow},
  \citenamefont {Luo},\ and\ \citenamefont {Mao}}]{Ding}%
  \BibitemOpen
  \bibfield  {author} {\bibinfo {author} {\bibfnamefont {Y.}~\bibnamefont
  {Ding}}, \bibinfo {author} {\bibfnamefont {R.}~\bibnamefont {Ahuja}},
  \bibinfo {author} {\bibfnamefont {J.}~\bibnamefont {Shu}}, \bibinfo {author}
  {\bibfnamefont {P.}~\bibnamefont {Chow}}, \bibinfo {author} {\bibfnamefont
  {W.}~\bibnamefont {Luo}}, \ and\ \bibinfo {author} {\bibfnamefont {H.-k.}\
  \bibnamefont {Mao}},\ }\href@noop {} {\bibfield  {journal} {\bibinfo
  {journal} {Phys. Rev. Lett.}\ }\textbf {\bibinfo {volume} {98}},\ \bibinfo
  {pages} {085502} (\bibinfo {year} {2007})}\BibitemShut {NoStop}%
\bibitem [{\citenamefont {Jenei}\ \emph {et~al.}(2011)\citenamefont {Jenei},
  \citenamefont {Liermann}, \citenamefont {Cynn}, \citenamefont {Klepeis},
  \citenamefont {Baer},\ and\ \citenamefont {Evans}}]{jenei}%
  \BibitemOpen
  \bibfield  {author} {\bibinfo {author} {\bibfnamefont {Z.}~\bibnamefont
  {Jenei}}, \bibinfo {author} {\bibfnamefont {H.~P.}\ \bibnamefont {Liermann}},
  \bibinfo {author} {\bibfnamefont {H.}~\bibnamefont {Cynn}}, \bibinfo {author}
  {\bibfnamefont {J.-H.~P.}\ \bibnamefont {Klepeis}}, \bibinfo {author}
  {\bibfnamefont {B.~J.}\ \bibnamefont {Baer}}, \ and\ \bibinfo {author}
  {\bibfnamefont {W.~J.}\ \bibnamefont {Evans}},\ }\href@noop {} {\bibfield
  {journal} {\bibinfo  {journal} {Phys. Rev. B}\ }\textbf {\bibinfo {volume}
  {83}},\ \bibinfo {pages} {054101} (\bibinfo {year} {2011})}\BibitemShut
  {NoStop}%
\bibitem [{\citenamefont {Yuying}\ \emph {et~al.}(2014)\citenamefont {Yuying},
  \citenamefont {Ye}, \citenamefont {Chengda}, \citenamefont {Xuemei},
  \citenamefont {Yinghua}, \citenamefont {Qiang},\ and\ \citenamefont
  {Hua}}]{vshock}%
  \BibitemOpen
  \bibfield  {author} {\bibinfo {author} {\bibfnamefont {Y.}~\bibnamefont
  {Yuying}}, \bibinfo {author} {\bibfnamefont {T.}~\bibnamefont {Ye}}, \bibinfo
  {author} {\bibfnamefont {D.}~\bibnamefont {Chengda}}, \bibinfo {author}
  {\bibfnamefont {L.}~\bibnamefont {Xuemei}}, \bibinfo {author} {\bibfnamefont
  {L.}~\bibnamefont {Yinghua}}, \bibinfo {author} {\bibfnamefont
  {W.}~\bibnamefont {Qiang}}, \ and\ \bibinfo {author} {\bibfnamefont
  {T.}~\bibnamefont {Hua}},\ }\href@noop {} {\bibfield  {journal} {\bibinfo
  {journal} {Appl. Phys. Lett.}\ }\textbf {\bibinfo {volume} {105}},\ \bibinfo
  {pages} {201910} (\bibinfo {year} {2014})}\BibitemShut {NoStop}%
\bibitem [{\citenamefont {Errandonea}\ \emph {et~al.}(2019)\citenamefont
  {Errandonea}, \citenamefont {MacLeod}, \citenamefont {Burakovsky},
  \citenamefont {Santamaria-Perez}, \citenamefont {Proctor}, \citenamefont
  {Cynn},\ and\ \citenamefont {Mezouar}}]{vphase}%
  \BibitemOpen
  \bibfield  {author} {\bibinfo {author} {\bibfnamefont {D.}~\bibnamefont
  {Errandonea}}, \bibinfo {author} {\bibfnamefont {S.~G.}\ \bibnamefont
  {MacLeod}}, \bibinfo {author} {\bibfnamefont {L.}~\bibnamefont {Burakovsky}},
  \bibinfo {author} {\bibfnamefont {D.}~\bibnamefont {Santamaria-Perez}},
  \bibinfo {author} {\bibfnamefont {J.~E.}\ \bibnamefont {Proctor}}, \bibinfo
  {author} {\bibfnamefont {H.}~\bibnamefont {Cynn}}, \ and\ \bibinfo {author}
  {\bibfnamefont {M.}~\bibnamefont {Mezouar}},\ }\href@noop {} {\bibfield
  {journal} {\bibinfo  {journal} {Phys. Rev. B}\ }\textbf {\bibinfo {volume}
  {100}},\ \bibinfo {pages} {094111} (\bibinfo {year} {2019})}\BibitemShut
  {NoStop}%
\bibitem [{\citenamefont {Wang}\ \emph {et~al.}(2021)\citenamefont {Wang},
  \citenamefont {Li}, \citenamefont {Zhou}, \citenamefont {Tan}, \citenamefont
  {Hao}, \citenamefont {Yu}, \citenamefont {Dai}, \citenamefont {Jin},
  \citenamefont {Wu}, \citenamefont {Jing}, \citenamefont {Chen}, \citenamefont
  {Yan}, \citenamefont {Wang},\ and\ \citenamefont {Geng}}]{wang2021}%
  \BibitemOpen
  \bibfield  {author} {\bibinfo {author} {\bibfnamefont {H.}~\bibnamefont
  {Wang}}, \bibinfo {author} {\bibfnamefont {J.}~\bibnamefont {Li}}, \bibinfo
  {author} {\bibfnamefont {X.~M.}\ \bibnamefont {Zhou}}, \bibinfo {author}
  {\bibfnamefont {Y.}~\bibnamefont {Tan}}, \bibinfo {author} {\bibfnamefont
  {L.}~\bibnamefont {Hao}}, \bibinfo {author} {\bibfnamefont {Y.~Y.}\
  \bibnamefont {Yu}}, \bibinfo {author} {\bibfnamefont {C.~D.}\ \bibnamefont
  {Dai}}, \bibinfo {author} {\bibfnamefont {K.}~\bibnamefont {Jin}}, \bibinfo
  {author} {\bibfnamefont {Q.}~\bibnamefont {Wu}}, \bibinfo {author}
  {\bibfnamefont {Q.~M.}\ \bibnamefont {Jing}}, \bibinfo {author}
  {\bibfnamefont {X.~R.}\ \bibnamefont {Chen}}, \bibinfo {author}
  {\bibfnamefont {X.~Z.}\ \bibnamefont {Yan}}, \bibinfo {author} {\bibfnamefont
  {Y.~X.}\ \bibnamefont {Wang}}, \ and\ \bibinfo {author} {\bibfnamefont
  {H.~Y.}\ \bibnamefont {Geng}},\ }\href@noop {} {\bibfield  {journal}
  {\bibinfo  {journal} {Phys. Rev. B}\ }\textbf {\bibinfo {volume} {104}},\
  \bibinfo {pages} {134102} (\bibinfo {year} {2021})}\BibitemShut {NoStop}%
\bibitem [{\citenamefont {Daniele}\ \emph {et~al.}(2016)\citenamefont
  {Daniele}, \citenamefont {Daniel~L.}, \citenamefont {Alexei}, \citenamefont
  {Chantel~M.}, \citenamefont {David~G.},\ and\ \citenamefont
  {Michael}}]{vexp2016}%
  \BibitemOpen
  \bibfield  {author} {\bibinfo {author} {\bibfnamefont {A.}~\bibnamefont
  {Daniele}}, \bibinfo {author} {\bibfnamefont {F.}~\bibnamefont {Daniel~L.}},
  \bibinfo {author} {\bibfnamefont {B.}~\bibnamefont {Alexei}}, \bibinfo
  {author} {\bibfnamefont {A.}~\bibnamefont {Chantel~M.}}, \bibinfo {author}
  {\bibfnamefont {R.}~\bibnamefont {David~G.}}, \ and\ \bibinfo {author}
  {\bibfnamefont {K.}~\bibnamefont {Michael}},\ }\href@noop {} {\bibfield
  {journal} {\bibinfo  {journal} {Sci. Rep.}\ }\textbf {\bibinfo {volume}
  {6}},\ \bibinfo {pages} {31887} (\bibinfo {year} {2016})}\BibitemShut
  {NoStop}%
\bibitem [{\citenamefont {Yi~X.}\ \emph {et~al.}(2016)\citenamefont {Yi~X.},
  \citenamefont {Q.}, \citenamefont {Xiang~R.},\ and\ \citenamefont
  {Hua~Y.}}]{wuqiang16}%
  \BibitemOpen
  \bibfield  {author} {\bibinfo {author} {\bibfnamefont {W.}~\bibnamefont
  {Yi~X.}}, \bibinfo {author} {\bibfnamefont {W.}~\bibnamefont {Q.}}, \bibinfo
  {author} {\bibfnamefont {C.}~\bibnamefont {Xiang~R.}}, \ and\ \bibinfo
  {author} {\bibfnamefont {G.}~\bibnamefont {Hua~Y.}},\ }\href@noop {}
  {\bibfield  {journal} {\bibinfo  {journal} {Sci. Rep.}\ }\textbf {\bibinfo
  {volume} {6}},\ \bibinfo {pages} {32419} (\bibinfo {year}
  {2016})}\BibitemShut {NoStop}%
\bibitem [{\citenamefont {Qiu}\ and\ \citenamefont {Marcus}(2008)}]{qiu2008}%
  \BibitemOpen
  \bibfield  {author} {\bibinfo {author} {\bibfnamefont {S.}~\bibnamefont
  {Qiu}}\ and\ \bibinfo {author} {\bibfnamefont {P.}~\bibnamefont {Marcus}},\
  }\href@noop {} {\bibfield  {journal} {\bibinfo  {journal} {J. Phys.: Condens.
  Matter}\ }\textbf {\bibinfo {volume} {20}},\ \bibinfo {pages} {275218}
  (\bibinfo {year} {2008})}\BibitemShut {NoStop}%
\bibitem [{\citenamefont {Luo}\ \emph {et~al.}(2007)\citenamefont {Luo},
  \citenamefont {Ahuja}, \citenamefont {Ding},\ and\ \citenamefont
  {Mao}}]{luo07}%
  \BibitemOpen
  \bibfield  {author} {\bibinfo {author} {\bibfnamefont {W.}~\bibnamefont
  {Luo}}, \bibinfo {author} {\bibfnamefont {R.}~\bibnamefont {Ahuja}}, \bibinfo
  {author} {\bibfnamefont {Y.}~\bibnamefont {Ding}}, \ and\ \bibinfo {author}
  {\bibfnamefont {H.-k.}\ \bibnamefont {Mao}},\ }\href@noop {} {\bibfield
  {journal} {\bibinfo  {journal} {Proc. Natl Acad. Sci.}\ }\textbf {\bibinfo
  {volume} {104}},\ \bibinfo {pages} {16428} (\bibinfo {year}
  {2007})}\BibitemShut {NoStop}%
\bibitem [{\citenamefont {A.~K.}\ and\ \citenamefont {P.}(2007)}]{Verma2007}%
  \BibitemOpen
  \bibfield  {author} {\bibinfo {author} {\bibfnamefont {V.}~\bibnamefont
  {A.~K.}}\ and\ \bibinfo {author} {\bibfnamefont {M.}~\bibnamefont {P.}},\
  }\href@noop {} {\bibfield  {journal} {\bibinfo  {journal} {Europhysics
  Letters}\ }\textbf {\bibinfo {volume} {81}},\ \bibinfo {pages} {37003}
  (\bibinfo {year} {2007})}\BibitemShut {NoStop}%
\bibitem [{\citenamefont {Lee}\ \emph {et~al.}(2007)\citenamefont {Lee},
  \citenamefont {Rudd}, \citenamefont {Klepeis}, \citenamefont {S\"oderlind},\
  and\ \citenamefont {Landa}}]{lee1}%
  \BibitemOpen
  \bibfield  {author} {\bibinfo {author} {\bibfnamefont {B.}~\bibnamefont
  {Lee}}, \bibinfo {author} {\bibfnamefont {R.~E.}\ \bibnamefont {Rudd}},
  \bibinfo {author} {\bibfnamefont {J.~E.}\ \bibnamefont {Klepeis}}, \bibinfo
  {author} {\bibfnamefont {P.}~\bibnamefont {S\"oderlind}}, \ and\ \bibinfo
  {author} {\bibfnamefont {A.}~\bibnamefont {Landa}},\ }\href@noop {}
  {\bibfield  {journal} {\bibinfo  {journal} {Phys. Rev. B}\ }\textbf {\bibinfo
  {volume} {75}},\ \bibinfo {pages} {180101} (\bibinfo {year}
  {2007})}\BibitemShut {NoStop}%
\bibitem [{\citenamefont {Lee}\ \emph {et~al.}(2008)\citenamefont {Lee},
  \citenamefont {Rudd}, \citenamefont {Klepeis},\ and\ \citenamefont
  {Becker}}]{lee2}%
  \BibitemOpen
  \bibfield  {author} {\bibinfo {author} {\bibfnamefont {B.}~\bibnamefont
  {Lee}}, \bibinfo {author} {\bibfnamefont {R.~E.}\ \bibnamefont {Rudd}},
  \bibinfo {author} {\bibfnamefont {J.~E.}\ \bibnamefont {Klepeis}}, \ and\
  \bibinfo {author} {\bibfnamefont {R.}~\bibnamefont {Becker}},\ }\href@noop {}
  {\bibfield  {journal} {\bibinfo  {journal} {Phys. Rev. B}\ }\textbf {\bibinfo
  {volume} {77}},\ \bibinfo {pages} {134105} (\bibinfo {year}
  {2008})}\BibitemShut {NoStop}%
\bibitem [{\citenamefont {Landa}\ \emph {et~al.}(2018)\citenamefont {Landa},
  \citenamefont {Söderlind}, \citenamefont {Naumov}, \citenamefont {Klepeis},\
  and\ \citenamefont {Vitos}}]{kohnanomaly2018}%
  \BibitemOpen
  \bibfield  {author} {\bibinfo {author} {\bibfnamefont {A.}~\bibnamefont
  {Landa}}, \bibinfo {author} {\bibfnamefont {P.}~\bibnamefont {Söderlind}},
  \bibinfo {author} {\bibfnamefont {I.~I.}\ \bibnamefont {Naumov}}, \bibinfo
  {author} {\bibfnamefont {J.~E.}\ \bibnamefont {Klepeis}}, \ and\ \bibinfo
  {author} {\bibfnamefont {L.}~\bibnamefont {Vitos}},\ }\href@noop {}
  {\bibfield  {journal} {\bibinfo  {journal} {Computation}\ }\textbf {\bibinfo
  {volume} {6}},\ \bibinfo {pages} {29} (\bibinfo {year} {2018})}\BibitemShut
  {NoStop}%
\bibitem [{\citenamefont {Akahama}\ \emph {et~al.}(2021)\citenamefont
  {Akahama}, \citenamefont {Kawaguchi}, \citenamefont {Hirao},\ and\
  \citenamefont {Ohishi}}]{akahama2021}%
  \BibitemOpen
  \bibfield  {author} {\bibinfo {author} {\bibfnamefont {Y.}~\bibnamefont
  {Akahama}}, \bibinfo {author} {\bibfnamefont {S.}~\bibnamefont {Kawaguchi}},
  \bibinfo {author} {\bibfnamefont {N.}~\bibnamefont {Hirao}}, \ and\ \bibinfo
  {author} {\bibfnamefont {Y.}~\bibnamefont {Ohishi}},\ }\href {\doibase
  10.1063/5.0041208} {\bibfield  {journal} {\bibinfo  {journal} {J. Appl.
  Phys.}\ }\textbf {\bibinfo {volume} {129}},\ \bibinfo {pages} {135902}
  (\bibinfo {year} {2021})}\BibitemShut {NoStop}%
\bibitem [{\citenamefont {Stevenson}\ \emph {et~al.}(2021)\citenamefont
  {Stevenson}, \citenamefont {Pace}, \citenamefont {Storm}, \citenamefont
  {Finnegan}, \citenamefont {Garbarino}, \citenamefont {Wilson}, \citenamefont
  {McGonegle}, \citenamefont {Macleod},\ and\ \citenamefont
  {McMahon}}]{mcmahon2021}%
  \BibitemOpen
  \bibfield  {author} {\bibinfo {author} {\bibfnamefont {M.~G.}\ \bibnamefont
  {Stevenson}}, \bibinfo {author} {\bibfnamefont {E.~J.}\ \bibnamefont {Pace}},
  \bibinfo {author} {\bibfnamefont {C.~V.}\ \bibnamefont {Storm}}, \bibinfo
  {author} {\bibfnamefont {S.~E.}\ \bibnamefont {Finnegan}}, \bibinfo {author}
  {\bibfnamefont {G.}~\bibnamefont {Garbarino}}, \bibinfo {author}
  {\bibfnamefont {C.~W.}\ \bibnamefont {Wilson}}, \bibinfo {author}
  {\bibfnamefont {D.}~\bibnamefont {McGonegle}}, \bibinfo {author}
  {\bibfnamefont {S.~G.}\ \bibnamefont {Macleod}}, \ and\ \bibinfo {author}
  {\bibfnamefont {M.~I.}\ \bibnamefont {McMahon}},\ }\href@noop {} {\bibfield
  {journal} {\bibinfo  {journal} {Phys. Rev. B}\ }\textbf {\bibinfo {volume}
  {103}},\ \bibinfo {pages} {134103} (\bibinfo {year} {2021})}\BibitemShut
  {NoStop}%
\bibitem [{\citenamefont {Ushcats}\ \emph {et~al.}(2016)\citenamefont
  {Ushcats}, \citenamefont {Bulavin}, \citenamefont {Sysoev}, \citenamefont
  {Bardik},\ and\ \citenamefont {Alekseev}}]{pfconden}%
  \BibitemOpen
  \bibfield  {author} {\bibinfo {author} {\bibfnamefont {M.~V.}\ \bibnamefont
  {Ushcats}}, \bibinfo {author} {\bibfnamefont {L.~A.}\ \bibnamefont
  {Bulavin}}, \bibinfo {author} {\bibfnamefont {V.~M.}\ \bibnamefont {Sysoev}},
  \bibinfo {author} {\bibfnamefont {V.~Y.}\ \bibnamefont {Bardik}}, \ and\
  \bibinfo {author} {\bibfnamefont {A.~N.}\ \bibnamefont {Alekseev}},\
  }\href@noop {} {\bibfield  {journal} {\bibinfo  {journal} {J. Mol. Liq.}\
  }\textbf {\bibinfo {volume} {224}},\ \bibinfo {pages} {694 } (\bibinfo {year}
  {2016})}\BibitemShut {NoStop}%
\bibitem [{\citenamefont {Martynov}(1999)}]{mce}%
  \BibitemOpen
  \bibfield  {author} {\bibinfo {author} {\bibfnamefont {G.~A.}\ \bibnamefont
  {Martynov}},\ }\href@noop {} {\bibfield  {journal} {\bibinfo  {journal}
  {Phys. Usp.}\ }\textbf {\bibinfo {volume} {42}},\ \bibinfo {pages} {517}
  (\bibinfo {year} {1999})}\BibitemShut {NoStop}%
\bibitem [{\citenamefont {Hansen}\ and\ \citenamefont {van
  Gunsteren}(2014)}]{fe}%
  \BibitemOpen
  \bibfield  {author} {\bibinfo {author} {\bibfnamefont {N.}~\bibnamefont
  {Hansen}}\ and\ \bibinfo {author} {\bibfnamefont {W.~F.}\ \bibnamefont {van
  Gunsteren}},\ }\href@noop {} {\bibfield  {journal} {\bibinfo  {journal} {J.
  Chem. Theory Comput.}\ }\textbf {\bibinfo {volume} {10}},\ \bibinfo {pages}
  {2632} (\bibinfo {year} {2014})}\BibitemShut {NoStop}%
\bibitem [{\citenamefont {Liu}(2020)}]{liu2020}%
  \BibitemOpen
  \bibfield  {author} {\bibinfo {author} {\bibfnamefont {Z.-K.}\ \bibnamefont
  {Liu}},\ }\href@noop {} {\bibfield  {journal} {\bibinfo  {journal} {Acta
  Materialia}\ }\textbf {\bibinfo {volume} {200}},\ \bibinfo {pages} {745}
  (\bibinfo {year} {2020})}\BibitemShut {NoStop}%
\bibitem [{\citenamefont {Ning}\ \emph {et~al.}(2021)\citenamefont {Ning},
  \citenamefont {Gong}, \citenamefont {Weng},\ and\ \citenamefont
  {Ning}}]{nby}%
  \BibitemOpen
  \bibfield  {author} {\bibinfo {author} {\bibfnamefont {B.-Y.}\ \bibnamefont
  {Ning}}, \bibinfo {author} {\bibfnamefont {L.-C.}\ \bibnamefont {Gong}},
  \bibinfo {author} {\bibfnamefont {T.-C.}\ \bibnamefont {Weng}}, \ and\
  \bibinfo {author} {\bibfnamefont {X.-J.}\ \bibnamefont {Ning}},\ }\href@noop
  {} {\bibfield  {journal} {\bibinfo  {journal} {J. Phys.: Condens. Matter}\
  }\textbf {\bibinfo {volume} {33}},\ \bibinfo {pages} {115901} (\bibinfo
  {year} {2021})}\BibitemShut {NoStop}%
\bibitem [{\citenamefont {Liu}\ \emph {et~al.}(2019)\citenamefont {Liu},
  \citenamefont {Ning}, \citenamefont {Gong}, \citenamefont {Weng},\ and\
  \citenamefont {Ning}}]{lyp}%
  \BibitemOpen
  \bibfield  {author} {\bibinfo {author} {\bibfnamefont {Y.-P.}\ \bibnamefont
  {Liu}}, \bibinfo {author} {\bibfnamefont {B.-Y.}\ \bibnamefont {Ning}},
  \bibinfo {author} {\bibfnamefont {L.-C.}\ \bibnamefont {Gong}}, \bibinfo
  {author} {\bibfnamefont {T.-C.}\ \bibnamefont {Weng}}, \ and\ \bibinfo
  {author} {\bibfnamefont {X.-J.}\ \bibnamefont {Ning}},\ }\href@noop {}
  {\bibfield  {journal} {\bibinfo  {journal} {Nanomaterials}\ }\textbf
  {\bibinfo {volume} {9}},\ \bibinfo {pages} {978} (\bibinfo {year}
  {2019})}\BibitemShut {NoStop}%
\bibitem [{\citenamefont {Gong}\ \emph {et~al.}(2019)\citenamefont {Gong},
  \citenamefont {Ning}, \citenamefont {Weng},\ and\ \citenamefont
  {Ning}}]{glc1}%
  \BibitemOpen
  \bibfield  {author} {\bibinfo {author} {\bibfnamefont {L.-C.}\ \bibnamefont
  {Gong}}, \bibinfo {author} {\bibfnamefont {B.-Y.}\ \bibnamefont {Ning}},
  \bibinfo {author} {\bibfnamefont {T.-C.}\ \bibnamefont {Weng}}, \ and\
  \bibinfo {author} {\bibfnamefont {X.-J.}\ \bibnamefont {Ning}},\ }\href@noop
  {} {\bibfield  {journal} {\bibinfo  {journal} {Entropy}\ }\textbf {\bibinfo
  {volume} {21}},\ \bibinfo {pages} {1050} (\bibinfo {year}
  {2019})}\BibitemShut {NoStop}%
\bibitem [{\citenamefont {Gong}\ \emph {et~al.}(2020)\citenamefont {Gong},
  \citenamefont {Ning}, \citenamefont {Ming}, \citenamefont {Weng},\ and\
  \citenamefont {Ning}}]{glc2}%
  \BibitemOpen
  \bibfield  {author} {\bibinfo {author} {\bibfnamefont {L.-C.}\ \bibnamefont
  {Gong}}, \bibinfo {author} {\bibfnamefont {B.-Y.}\ \bibnamefont {Ning}},
  \bibinfo {author} {\bibfnamefont {C.}~\bibnamefont {Ming}}, \bibinfo {author}
  {\bibfnamefont {T.-C.}\ \bibnamefont {Weng}}, \ and\ \bibinfo {author}
  {\bibfnamefont {X.-J.}\ \bibnamefont {Ning}},\ }\href@noop {} {\bibfield
  {journal} {\bibinfo  {journal} {J. Phys.: Condens. Matter}\ }\textbf
  {\bibinfo {volume} {33}},\ \bibinfo {pages} {085901} (\bibinfo {year}
  {2020})}\BibitemShut {NoStop}%
\bibitem [{\citenamefont {Dierckx}(1975)}]{spl1}%
  \BibitemOpen
  \bibfield  {author} {\bibinfo {author} {\bibfnamefont {P.}~\bibnamefont
  {Dierckx}},\ }\href {\doibase https://doi.org/10.1016/0771-050X(75)90034-0}
  {\bibfield  {journal} {\bibinfo  {journal} {Journal of Computational and
  Applied Mathematics}\ }\textbf {\bibinfo {volume} {1}},\ \bibinfo {pages}
  {165} (\bibinfo {year} {1975})}\BibitemShut {NoStop}%
\bibitem [{\citenamefont {Dierckx}(1985)}]{spl2}%
  \BibitemOpen
  \bibfield  {author} {\bibinfo {author} {\bibfnamefont {P.}~\bibnamefont
  {Dierckx}},\ }\href@noop {} {\bibfield  {journal} {\bibinfo  {journal} {SIAM
  J.Numer.Anal}\ }\textbf {\bibinfo {volume} {19}},\ \bibinfo {pages} {1286}
  (\bibinfo {year} {1985})}\BibitemShut {NoStop}%
\bibitem [{\citenamefont {Kresse}\ and\ \citenamefont
  {Furthm\"uller}(1996{\natexlab{a}})}]{VASP1}%
  \BibitemOpen
  \bibfield  {author} {\bibinfo {author} {\bibfnamefont {G.}~\bibnamefont
  {Kresse}}\ and\ \bibinfo {author} {\bibfnamefont {J.}~\bibnamefont
  {Furthm\"uller}},\ }\href@noop {} {\bibfield  {journal} {\bibinfo  {journal}
  {Comput. Mat. Sci.}\ }\textbf {\bibinfo {volume} {6}},\ \bibinfo {pages} {15}
  (\bibinfo {year} {1996}{\natexlab{a}})}\BibitemShut {NoStop}%
\bibitem [{\citenamefont {Kresse}\ and\ \citenamefont
  {Furthm\"uller}(1996{\natexlab{b}})}]{VASP2}%
  \BibitemOpen
  \bibfield  {author} {\bibinfo {author} {\bibfnamefont {G.}~\bibnamefont
  {Kresse}}\ and\ \bibinfo {author} {\bibfnamefont {J.}~\bibnamefont
  {Furthm\"uller}},\ }\href@noop {} {\bibfield  {journal} {\bibinfo  {journal}
  {Phys. Rev. B}\ }\textbf {\bibinfo {volume} {54}},\ \bibinfo {pages} {11169}
  (\bibinfo {year} {1996}{\natexlab{b}})}\BibitemShut {NoStop}%
\bibitem [{\citenamefont {Bl\"ochl}(1994)}]{paw1}%
  \BibitemOpen
  \bibfield  {author} {\bibinfo {author} {\bibfnamefont {P.~E.}\ \bibnamefont
  {Bl\"ochl}},\ }\href@noop {} {\bibfield  {journal} {\bibinfo  {journal}
  {Phys. Rev. B}\ }\textbf {\bibinfo {volume} {50}},\ \bibinfo {pages} {17953}
  (\bibinfo {year} {1994})}\BibitemShut {NoStop}%
\bibitem [{\citenamefont {Kresse}\ and\ \citenamefont {Joubert}(1999)}]{paw2}%
  \BibitemOpen
  \bibfield  {author} {\bibinfo {author} {\bibfnamefont {G.}~\bibnamefont
  {Kresse}}\ and\ \bibinfo {author} {\bibfnamefont {D.}~\bibnamefont
  {Joubert}},\ }\href@noop {} {\bibfield  {journal} {\bibinfo  {journal} {Phys.
  Rev. B}\ }\textbf {\bibinfo {volume} {59}},\ \bibinfo {pages} {1758}
  (\bibinfo {year} {1999})}\BibitemShut {NoStop}%
\bibitem [{\citenamefont {Perdew}\ \emph {et~al.}(1996)\citenamefont {Perdew},
  \citenamefont {Burke},\ and\ \citenamefont {Ernzerhof}}]{pbe}%
  \BibitemOpen
  \bibfield  {author} {\bibinfo {author} {\bibfnamefont {J.~P.}\ \bibnamefont
  {Perdew}}, \bibinfo {author} {\bibfnamefont {K.}~\bibnamefont {Burke}}, \
  and\ \bibinfo {author} {\bibfnamefont {M.}~\bibnamefont {Ernzerhof}},\
  }\href@noop {} {\bibfield  {journal} {\bibinfo  {journal} {Phys. Rev. Lett.}\
  }\textbf {\bibinfo {volume} {77}},\ \bibinfo {pages} {3865} (\bibinfo {year}
  {1996})}\BibitemShut {NoStop}%
\bibitem [{\citenamefont {Zhang}\ \emph {et~al.}(2021)\citenamefont {Zhang},
  \citenamefont {Wang}, \citenamefont {Xian}, \citenamefont {Wang},
  \citenamefont {Fang}, \citenamefont {Duan}, \citenamefont {Gao},
  \citenamefont {Song},\ and\ \citenamefont {Liu}}]{gga1}%
  \BibitemOpen
  \bibfield  {author} {\bibinfo {author} {\bibfnamefont {T.}~\bibnamefont
  {Zhang}}, \bibinfo {author} {\bibfnamefont {Y.}~\bibnamefont {Wang}},
  \bibinfo {author} {\bibfnamefont {J.}~\bibnamefont {Xian}}, \bibinfo {author}
  {\bibfnamefont {S.}~\bibnamefont {Wang}}, \bibinfo {author} {\bibfnamefont
  {J.}~\bibnamefont {Fang}}, \bibinfo {author} {\bibfnamefont {S.}~\bibnamefont
  {Duan}}, \bibinfo {author} {\bibfnamefont {X.}~\bibnamefont {Gao}}, \bibinfo
  {author} {\bibfnamefont {H.}~\bibnamefont {Song}}, \ and\ \bibinfo {author}
  {\bibfnamefont {H.}~\bibnamefont {Liu}},\ }\href {\doibase 10.1063/5.0059360}
  {\bibfield  {journal} {\bibinfo  {journal} {Matter Radiat. Extremes}\
  }\textbf {\bibinfo {volume} {6}},\ \bibinfo {pages} {068401} (\bibinfo {year}
  {2021})}\BibitemShut {NoStop}%
\bibitem [{\citenamefont {Monkhorst}\ and\ \citenamefont
  {Pack}(1976)}]{monkhorst}%
  \BibitemOpen
  \bibfield  {author} {\bibinfo {author} {\bibfnamefont {H.~J.}\ \bibnamefont
  {Monkhorst}}\ and\ \bibinfo {author} {\bibfnamefont {J.~D.}\ \bibnamefont
  {Pack}},\ }\href@noop {} {\bibfield  {journal} {\bibinfo  {journal} {Phys.
  Rev. B}\ }\textbf {\bibinfo {volume} {13}},\ \bibinfo {pages} {5188}
  (\bibinfo {year} {1976})}\BibitemShut {NoStop}%
\end{thebibliography}
%

\end{document}